\documentclass[aps,twocolumn]{revtex4}
\usepackage{graphicx}
\usepackage{latexsym}
\usepackage{amsmath}
\usepackage{amsfonts}
\usepackage{amssymb}
\usepackage{braket}
\usepackage{xcolor}
\newcommand{\be}{\begin{equation}}
\newcommand{\ee}{\end{equation}}
\newcommand{\bea}{\begin{eqnarray}}
\newcommand{\eea}{\end{eqnarray}}
\begin{document}

\title{Evolutionary Advantage of Cell Size Control}
\date{\today}
\author{Spencer Hobson-Gutierrez$^{1}$ and Edo Kussell$^{1,2}$}
\affiliation{$^1$Department of Biology, New York University, 12 Waverly Place, New York, NY 10003. \\
$^2$Department of Physics, New York University, 726 Broadway, New York, NY 10003. }

\begin{abstract}
We analyze the advantage of cell size control strategies in growing populations under mortality constraints.  We demonstrate a general advantage of the adder control strategy in the presence of growth-dependent mortality, and for different size-dependent mortality landscapes.  Its advantage stems from epigenetic heritability of cell size, which enables selection to act on the distribution of cell sizes in a population to avoid mortality thresholds and adapt to a mortality landscape.
\end{abstract}
\maketitle
Cells of bacteria and other organisms grow exponentially in size before dividing \cite{Wang:2010aa,JunReview2018}. Noise in cell division timing is therefore expected to yield increasingly large variations in cell size over time, yet single cell lineages maintain a narrow distribution of cell sizes over many generations \cite{Campos2014-ql,Taheri-Araghi2015-am,Deforet2015,Susman2018-tw}.  While strategies \cite{Amir2014-li,Amir2017-py,Barber2021-ny,Francois2022} and molecular mechanisms \cite{Schmoller:2015aa,Si2019-rd,KCHuang_2020,doi:10.1073/pnas.2016391118} that underlie cell size control are being elucidated, it is presently unknown what selective advantage they confer in nature.  Specifically, in free-living microbes such as bacteria, if exponential biomass growth can be achieved by any number of cells, what evolutionary benefit does the control of cell size contribute to a growing population?  This question underlies our understanding of the evolutionary determinants of cell size and the origins of cell size control.

Starting from a single, exponentially growing cell that lacks a size control mechanism, the resulting population's average biomass growth rate would not depend strongly on how frequently the cells divide; rather, it would largely depend on the mean growth rate of the size of single cells.  However, if cells can die, and mortality is experienced independently by each cell, then each cell division halves the probability of extinction along a lineage.  By this reasoning, selection will favor lineages that divide as frequently as possible. Since cells cannot be too small due to biophysical constraints -- e.g. there must be enough room for their DNA and proteins -- the presence of mortality could select for cell size control.  In particular, we expect a control mechanism that enables cells to be as small as physically possible to be advantageous.  As the minimum viable size of cells is approached, errors in cell size control will become increasingly costly, as they can result in non-viable cells.  Natural selection should favor minimizing the errors and evolutionarily tuning the control mechanism to avoid such mortality thresholds.  This argument thus identifies a general advantage of tight cell size control.  Yet, different control strategies, e.g. `sizer' or `adder' \cite{lagomarsino_pnas_2014,Amir2014-li,amir_cellsystems_2017,Barber2021-ny} can efficiently minimize cell division errors, and it is therefore not clear whether one strategy is advantageous relative to another in a growing population that experiences mortality.

Here, we introduce a simple model that represents the above evolutionary forces. Using simulations, numerics, and analytical theory, we demonstrate that an adder strategy has a pronounced advantage over a sizer in the vicinity of mortality thresholds. We show that this advantage is due to epigenetic heritability of cell size, which is present in an adder strategy but absent in a sizer.  Cell size heritability enables within-population selection to act on the cell size distribution, moving it away from thresholds, and enhancing population growth in different mortality landscapes.

We first analyze the statistics of cell size control in isolated single cells, which can be observed experimentally in a `mother machine' device \cite{Susman2018-tw,Wang:2010aa}. Starting from a given birth size $l$, each cell grows exponentially in size and divides symmetrically. Given a parent cell's birth size $l$, the cell size control mechanism determines $\Delta$, the size added before cell division, a random variable whose fluctuations characterize the intrinsic noise of the control process (``input noise'').  The offspring's birth size $l'$ is determined by the \emph{size relation}, $l' = (l + \Delta)/2$.  The output of iterating this control process results in size fluctuations (``output noise''). In an \emph{adder mechanism}, cells attempt to add a fixed size before dividing, hence $\Delta$ and $l$ are independent random variables, while $l'$ and $l$ are positively correlated.  In this case, at steady-state the size relation implies that $\kappa_n(l) = \kappa_n(\Delta)/(2^n - 1)$, where $\kappa_n(X)$ denotes the $n$-th cumulant of random variable $X$.  In a \emph{sizer mechanism}, cells attempt to elongate to a fixed target size before dividing; hence $l'$ and $l$ are independent, while $\Delta$ ($=2l' - l$) is negatively correlated with $l$. At steady-state, the size relation implies $\kappa_n(l) = \kappa_n(\Delta)/[2^n + (-1)^n]$.  Such cumulant relations determine the input-output noise relation for each control mechanism \cite{supp}.  Considering the variance ($n = 2$), for the same input noise ($\sigma^2 \equiv \mathrm{Var}(\Delta)$) the output noise in the sizer ($\sigma^2 / 5$) is lower than that of the adder ($\sigma^2 / 3$), and in general $|\kappa_{n,sizer}(l)| \leq |\kappa_{n,adder}(l)|$ for fixed $\kappa_n(\Delta)$.  Thus, in isolated single cells, given the same intrinsic noise a sizer mechanism is better able to control birth size fluctuations than an adder mechanism.

\begin{figure}
\includegraphics[width=3.4in]{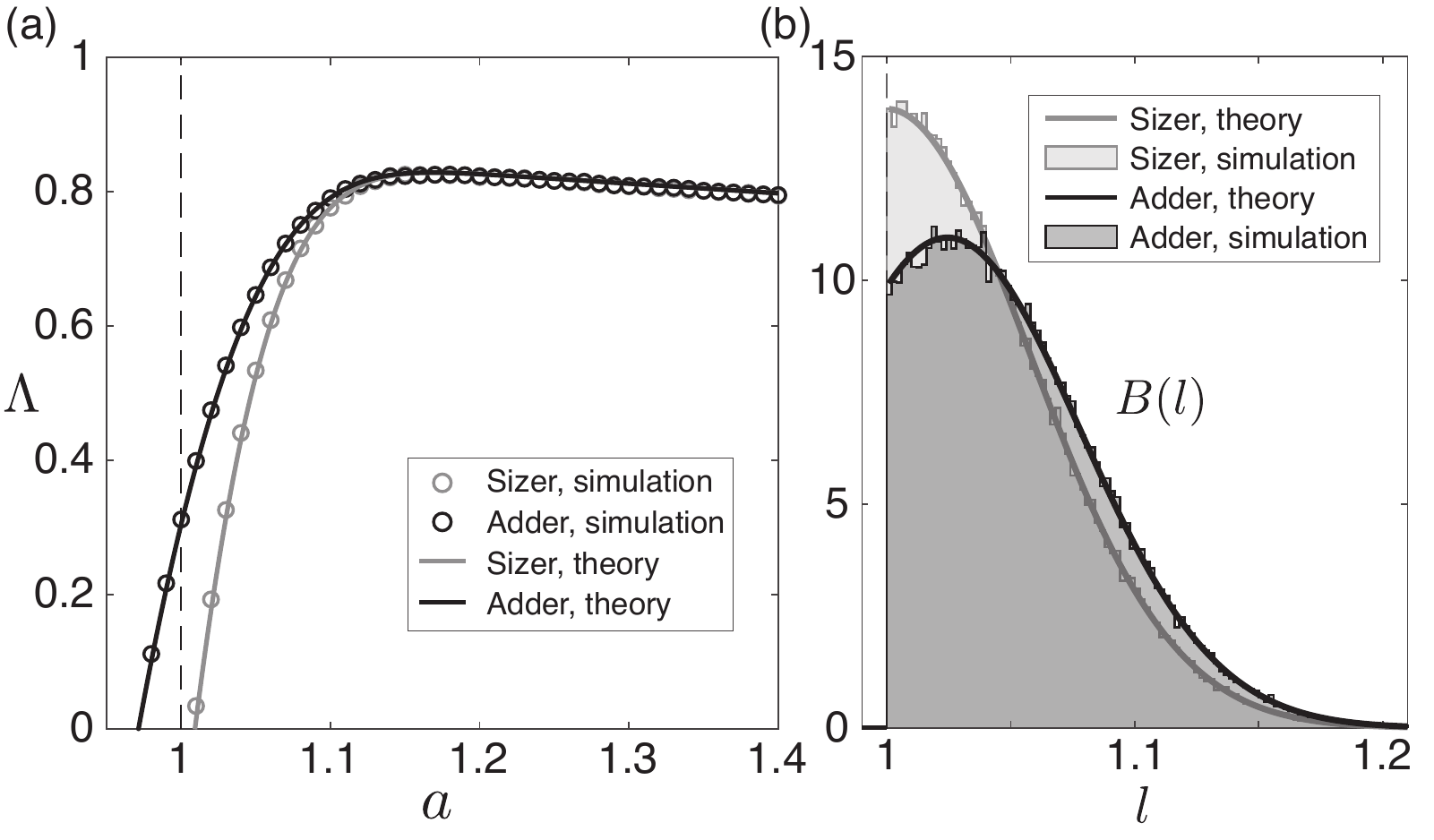}
\caption{Numerical and simulation results for population growth of adder and sizer mechanisms with mortality. (a) Growth rates measured in simulations (points) are compared with predictions (curves) from numerical solution of the transfer operator (Eq. \ref{eq:psi_steadystate}).  The vertical dashed line indicates $l_{min}$. Sizer and adder control mechanisms, $C_S(l')$ and $C_A(\Delta)$, are Gaussian with variances $\sigma^2_S$ and $\sigma_A^2$, respectively, and mean $a$. Parameter values are $\sigma_A = 0.1$, $\sigma_S = \sigma_A / \sqrt{3}$, $\beta = 0.1$, $l_{min} = 1$, and $\lambda_0 = 1$; the choice of $\sigma_S$ ensures that sizer and adder have identical output statistics in isolated cells. (b) Birth size distributions measured in simulation (shaded regions) and predicted by Eq. \ref{eq:psi_steadystate} (curves) for $a = l_{min}$; all other parameter values as in panel a.}
\label{fig:intro}
\end{figure}

We next consider a population of cells growing in the presence of mortality constraints. Active cell growth involves remodeling basic cellular structures, e.g. the cell wall of bacteria, and renders cells more susceptible to environmental stresses such as antibiotics \cite{Stokes:2019aa,doi:10.1128/mBio.02456-20}. Mortality in our model is determined by a constant $\beta$ characterizing growth-dependent mortality, where the probability of cell death during a growth increment $\delta l$ is given by $\beta \, \delta l$; and by a mortality threshold $l_{min}$, such that cells with birth size $l < l_{min}$ are not viable. In a sizer model, cell division occurs when size $2l'$ is reached, where $l'$ is a random variable with distribution $C_S(l')$.  In an adder model, division occurs when when the added size reaches $\Delta$, which is distributed according to $C_A(\Delta)$.  As cells grow and divide, the population size $N(t)$ increases exponentially, and its long-term growth rate is $\Lambda \equiv (1/t) \ln N(t)$, in the limit of large $t$. For sizer and adder mechanisms with matched output statistics in isolated cells (i.e. identical steady-state distribution of $l$), we compared the long-term growth rate of their populations as the target birth size $a \equiv \bar l$ was varied (Fig. 1a).  We found that for values of $a \gg l_{min}$, the two curves nearly overlap and decay linearly with slope $-\beta \lambda_0 / \ln 2$, where $\lambda_0$ is the exponential rate of cell size growth.  As $a$ approaches $l_{min}$, however, the curves separate and the adder mechanism exhibits a pronounced advantage over the sizer.  Strikingly, if we set the target birth size precisely at the mortality threshold, i.e. $a = l_{min}$, the sizer population does not grow, as half of its cells are born with $l' < l_{min}$, yet the adder population has a positive growth rate.  Moreover, the adder population can grow exponentially even for a range of target birth sizes $a < l_{min}$ over which the sizer population goes extinct.  This suggests that although the two size control mechanisms are equally precise, the structure of single cell lineages that they generate differs in a critical way.  Indeed, the distribution of birth sizes in the population with $a = l_{min}$ has a peak at $l = l_{min}$ for the sizer mechanism, but in the adder mechanism the peak is shifted to a value higher than $l_{min}$ (Fig. \ref{fig:intro}b).

To elucidate this phenomenon, we determined the lineage structure of sizer and adder populations using the transfer operator method for path integrals \cite{Feynman2010-ep,kussell_evolution_2012}.  A single cell lineage $\xi = (l_1, l_2, \ldots, l_n)$ is specified by a series of cell divisions $i = 1 \ldots n$, where the $i$-th division yields a cell with birth size $l_i$. We let $K(l', l)$ denote the expected number of offspring of size $l'$ for a parent of size $l$.  This kernel encapsulates both the cell size control strategy and the multiplicative fitness, including both cell division and mortality events; in the absence of mortality for binary cell division we have $\int K(l', l) dl' = 2$.  We let $B_t(l')$ denote the expected number of births in the population at time $t$ having birth size $l'$, and $\tilde B_t(l)$ the expected number of births at time $t$ that descend from a cell born with size $l$ at time zero. In other words, $\tilde B_t(l)$ counts the number of lineages in the population conditional on starting at $l$, while $B_t(l')$ counts lineages conditional on ending with $l'$.  Each birth with size $l'$ at time $t$ results from a parent cell of birth size $l$ that grew and divided at size $2l'$.  The parent itself was born at time $t - \tau$, where $\tau = (1 / \lambda) \ln (2l' / l)$, hence
\begin{equation}
B_t(l')  =   \int K(l', l) B_{t - \tau}(l) p(\lambda)  \, dl d\lambda \ , \label{eq:right_dynamics}
\end{equation}
where we assume that single-cell growth rate, $\lambda$, is independent of birth size and has probability distribution $p(\lambda)$. In rod-shaped bacteria such as \emph{E. coli}, which maintain a constant cell width, the birth size $l$ is measured by the cell length and $\lambda$ is known as the elongation rate; experiments indicate that $\lambda$ and $l$ are nearly uncorrelated \cite{lagomarsino_pnas_2014,JunReview2018}.

At steady-state population growth, $B_t(l) \simeq e^{\Lambda t} B(l)$, hence
\begin{equation}
B(l') =  \int  (2l' / l)^{-\Lambda/\lambda} K(l', l) B(l) p(\lambda)   \, dl d\lambda \ . \label{eq:B_steadystate}
\end{equation}

We first analyze the case of constant elongation rate $\lambda_0$, i.e. $p(\lambda) = \delta(\lambda - \lambda_0$) and then generalize to non-constant $\lambda$; we will see that the evolutionary advantage of cell size control is already apparent for the case of fixed $\lambda_0$. If we define $\psi(l) \equiv l^{\Lambda / \lambda_0} B(l)$, and $\alpha \equiv (\Lambda / \lambda_0) \ln 2$, we find
\begin{equation}
\psi(l') = e^{-\alpha} \int K(l', l)  \psi(l) dl \ , \label{eq:psi_steadystate}
\end{equation}
where $e^\alpha$ is the top eigenvalue of the kernel $K(l',l)$ and $\psi$ is the associated right eigenfunction corresponding to the `ground-state' \footnote{For the kernels that we study here, there is a unique physical solution corresponding to a positive eigenfunction with a real eigenvalue \cite{supp}.}, which determines the steady-state composition of newborn cells in the population via $B(l) = l^{-\Lambda/\lambda_0} \psi(l)$.  Analogously, we have $\tilde B(l) = l ^{\Lambda/\lambda_0} \tilde \psi(l)$, where $\tilde \psi$ is the corresponding left eigenfunction of the kernel $K$.  The density $\rho$ of states along lineages in the population is given by the usual expression from quantum mechanics, $\rho(l) \sim \tilde B(l) B(l) = \tilde \psi(l) \psi(l)$.

We now consider the specific forms of the kernel for sizer ($K_S$) and adder ($K_A$) mechanisms given by
\begin{align}
 K_S(l',l) &\equiv z C_S(l') e^{-\beta (2l'-l)} \theta(l' - l_{min}) \label{eq:sizer_kernel} \\
 K_A(l',l) &\equiv 2 z C_A(2l'-l) e^{-\beta (2l'-l)} \theta(l' - l_{min}) \label{eq:adder_kernel} \ ,
 \end{align}
where the growth-dependent mortality $\beta$ acts on the added size $\Delta = 2l' - l$, and the Heaviside function $\theta$ enforces the mortality threshold at $l_{min}$  \footnote{For the sizer kernel, an additional constraint may be necessary to ensure that the added size $2l' - l$ is positive. It is automatically satisfied for any distribution $C_S(l)$ whose support lies on an interval $[a,b]$, where $b < 2a$; alternatively, if the distribution is sufficiently narrow, the constraint can be neglected. We leave it out of the $K_S$ kernel which simplifies the analysis without impacting our results, as seen from the agreement with simulations in which the constraint is present. }. The constant $z$ is used to distinguish lineages within a population where each binary cell division generates two offspring (i.e. $z = 2$) from isolated single cell lineages ($z = 1$) \footnote{A value of $z < 2$ can be used to incorporate mortality that occurs at cell division.  Values of $z> 2$ correspond to mechanisms that can generate more than 2 offspring at each division, which changes the calculation of the added size and requires additional modification of the kernel.}.  The extra factor of $2$ in $K_A$ is due to the Jacobian of the transformation $\Delta \leftrightarrow 2l' - l$, which normalizes $C_A(2l' - l)$ when integrated over $l'$.  We verified that numerical solution of Eq. \ref{eq:psi_steadystate} for adder and sizer kernels correctly predicts the simulation results (Fig. 1).

To compare population growth of sizer and adder mechanisms, we first match their output statistics in isolated cells.  According to the cumulant relation, we must have $\ln \hat C_A(\omega) = \ln [\hat C_S(2\omega) / \hat C_S(\omega)]$, where $\hat C$ denotes the Laplace transform of $C_S(l)$ (for sizer) or $C_A(\Delta)$ (for adder), which ensures identical birth size distributions of the mechanisms \cite{supp}.  For a population growing under constant growth-dependent mortality ($\beta \geq 0$) with no threshold ($\l_{min} = 0$), solving Eq. \ref{eq:psi_steadystate} we obtain $\alpha = \ln z + \ln \hat C(\beta)$.  It is clear that in the absence of mortality ($\beta = 0$), $\alpha = \ln z$, and therefore $\Lambda = \lambda_0$, i.e. the population growth rate is equal to the elongation rate regardless of the details of the mechanism. In the presence of mortality ($\beta > 0$), the difference in growth rates between adder and sizer then satisfies $\alpha_A - \alpha_S =  \ln \hat C_S(2\beta) - 2 \ln \hat C_S(\beta)$, and a simple convexity argument \cite{supp} shows that $\alpha_A > \alpha_S$ holds in general \footnote{We note that $\alpha_A = \alpha_S$ only for noiseless size control, i.e. if  $C_S(l) = \delta(l - a)$ and $C_A(\Delta) = \delta(\Delta - a)$, which implies $\kappa_{n>1}(l) = 0$.}.  Therefore, for constant growth-dependent mortality, an adder mechanism has a long-term growth rate advantage over a sizer mechanism.

To study cell size control in the presence of a mortality threshold ($l_{min} > 0$), we specialize to Gaussian control via $C_A(\Delta) = g(\Delta)$ and $C_S(l) = g(l)$, where $g(x) \equiv (2 \pi \sigma^2)^{-1/2}e^{ - \frac{(x - a)^2}{2 \sigma^2}}$.  In this case, we know from the cumulant relation that both models have Gaussian output statistics in isolated cells, and by choosing $\sigma^2 = \sigma^2_A$ for the adder and $\sigma^2 = \sigma^2_S = \sigma^2_A / 3$ for sizer, we match their statistics. Birth sizes along a sizer lineage are independent; thus, the system behaves like a gas of particles in a potential well with no particle-particle interactions. In contrast, the form of the adder kernel (Eq. \ref{eq:adder_kernel}) is similar to that of a polymer, where the location $l'$ of the next monomer along the chain depends on the location $l$ of the previous monomer \footnote{The detailed form of the kernel, however, differs from that of a polymer, as the distance between adjacent monomers (i.e. parent-offspring pairs) is measured by $2l' - l$ rather than $l' - l$.}. 

The sizer kernel is separable, with $K_S = \psi(l') \tilde \psi(l)$, where $\psi(l') = z g(l') e^{-2 \beta l'} \theta(l' - l_{min})$ and $\tilde \psi(l) = e^{\beta l}$. The top eigenvalue is given by $\int \tilde \psi(l) \psi(l) dl$, from which we obtain
\begin{equation}
\alpha_S =-  \beta a + \frac{\beta^2 \sigma^2}{2} + \ln \left[ \frac{z}{2} \mathrm{erfc}\left(\frac{l_{min} - a + \beta \sigma^2}{\sigma \sqrt 2 } \right) \right] \label{eq:alpha_sizer} \ ,
\end{equation}
where $\mathrm{erfc}(x) \equiv \tfrac{2}{\sqrt{\pi}} \int_x^\infty  e^{-l^2} dl$ is the complementary error function.  The density of states is given by
\begin{equation}
\rho(l) \sim \tilde \psi(l) \psi(l) \sim e^{ - \frac{(l - a + \beta \sigma^2)^2}{2 \sigma^2}} \theta(l- l_{min}) \ , \label{eq:rho_sizer}
\end{equation}
indicating that lineages in the population shift the peak of the birth-size distribution to be centered at $a - \beta \sigma^2$.
\begin{figure}
\includegraphics[width=3.4in]{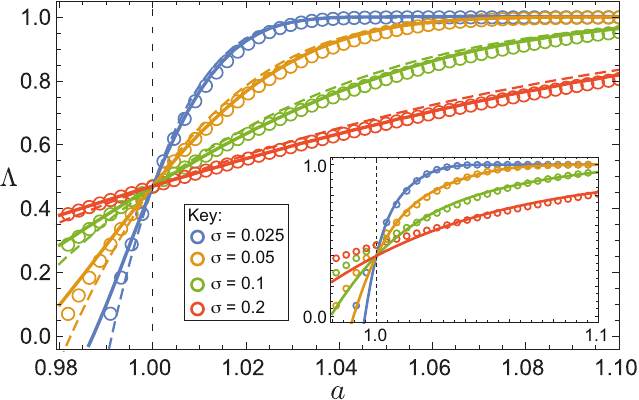}
\caption{Analytical results in the adder model.  Numerically computed values (points) and analytical approximations (curves) are shown for the long-term growth rate, $\Lambda$, as a function of the mean birth size, $a$.  RCS approximation, $r = 1.887$ (dashed curves) and FCS approximation, $h = 0.384$ (solid curves). Vertical dashed line indicates $l_{min}$.  Inset: Perturbation theory to the second order (solid line). Parameter values $\sigma$ are indicated in the key; additionally, $\beta = 0$, $\l_{min} = 1$, $z = 2$, and $\lambda = 1$.  See \cite{supp} for $\beta = 0.1$.}
\label{fig:approx}
\end{figure}
For $\beta = 0$, $\alpha_S$ is a monotonically increasing function of $a$ indicating that arbitrarily large cell sizes would be favored to avoid the mortality threshold.  For $\beta > 0$, there exists an optimal birth size $a_{opt}$ that maximizes $\alpha_S$ \cite{supp}, which is given by
\begin{equation}
a_{opt} \approx l_{min} + \sigma \sqrt{ \ln [1/ \left( 2 \pi \beta^{2} \sigma^{2} \right)]}
\end{equation}
in the limit of $\sigma \ll \beta^{-1}$, corresponding to tight cell size control \footnote{There may be other constraints on the parameters in different organisms, which may change the value of $a_{opt}$. For example, physiological growth laws may relate $\lambda_0$ and $a$ \cite{Taheri-Araghi2015-am,JunReview2018}. Such constraints can be incorporated when computing the long-term growth rate optimum via $\Lambda = \lambda_0 \alpha / \ln 2$.}.      The first-order effect of increasing noise $\sigma$ in size control is to move $a_{opt}$ away from the mortality threshold, decreasing the long-term growth rate (Eq. \ref{eq:alpha_sizer}).  

For the adder kernel, there does not appear to be an analytical solution, and we proceed using perturbative or scaling approximations. When $\sigma \ll a - l_{min}$, the kernel is strongly peaked around $a$ and the contribution of the mortality threshold at $l_{min}$ is small.  We can therefore rewrite the kernel as $K_A = K_0 + K_1$, where $K_0(l', l) \equiv 2 z g(2l' - l) e^{-\beta(2l' - l)}$, and $K_1 = K_A - K_0$.  The $K_0$ kernel corresponds to unconstrained integration over the whole space, including forbidden values $l' < l_{min}$, while the $K_1$ kernel accounts for these unphysical paths and removes them.  As long as $K_1 \ll K_0$, we can expand $\psi(l)$ using classical perturbation theory.  Perturbation expansion to the second order predicts the growth rate reasonably well for $a > l_{min}$ (Fig. \ref{fig:approx}, inset).  The theory correctly predicts positive growth at the mortality threshold, e.g. for $\beta = 0$ it predicts $\Lambda \approx 0.411$ at $a = l_{min}$, while the numerically computed value is 0.470 \cite{supp}.  It also shows that for $\beta = 0$ at $a = l_{min}$ the value of the growth rate is independent of the size of fluctuations $\sigma$ (Fig. \ref{fig:approx}, inset).  For $\beta > 0$, curves with different values of $\sigma$ converge near $a = l_{min}$ (see Fig. A2 in \cite{supp}).  This suggests that a scaling approach may be useful to understand the behavior near the boundary. While the second order perturbation theory is accurate for $a > l_{min}$, below the threshold larger deviations are apparent, as paths that cross the threshold become increasingly common and must be accounted for.  Although higher order perturbations can be computed, their analytical form is unwieldy, and we turn to a scaling-based approach below.

By analogy with polymers, we can model the lineages in the adder model as rigid rods with a certain persistence length.  To this end, we assume that offspring perfectly inherit their parent's birth size for $r$ generations.  Each such `block' of $r$ monomers corresponds to a single, random choice of birth size $l$, and the blocks are therefore independent just like the individual monomers in the sizer model; we therefore call this the `Rigid Correlated Sizer' (RCS) approximation.  The kernel on blocks of size $r$ is given by
\begin{equation}
 K_{RCS}(l',l) \equiv z^r g(l') e^{-\beta [(r+1)l'-l]}  \theta(l' - l_{min}) \ .
\end{equation}
We reinterpret $B_t(l')$ in Eq. \ref{eq:right_dynamics} as the birth rate of blocks with birth size $l'$ at $t$. These are offspring from parent blocks of birth size $l$ born at $t - \tau$, where $\tau = (1 / \lambda_0) \ln (2^r l' / l)$, which yields $\alpha = r (\Lambda / \lambda_0) \ln 2$ in Eq. \ref{eq:psi_steadystate}.  Computing the top eigenvalue of $K_{RCS}$, we obtain the same form as for the sizer kernel (Eq. \ref{eq:alpha_sizer}) with parameter rescalings $\beta \leftrightarrow r \beta$, $z \leftrightarrow z^r$, and $\alpha \leftrightarrow r \alpha$, yielding
\begin{equation}
\alpha_{RCS} =-  \beta a + \frac{r \beta^2 \sigma^2}{2} + \frac{1}{r} \ln \left[ \frac{z^r}{2} \mathrm{erfc}\left(\frac{l_{min} - a + r \beta \sigma^2}{\sigma \sqrt 2 } \right) \right] \ .
\end{equation}
To ensure that the RCS birth size distribution matches that of isolated single cells in the adder model, we substitute $\sigma^2 \rightarrow \sigma_A^2 / 3$ above.  This leaves only the parameter $r$ to be determined. The correlation of $l'$ and $l$ for isolated cells in the absence of mortality is equal to $1 - 1/r$; hence to match the correlation in the adder model it is reasonable to take $r = 2$.  Alternatively, at $a = l_{min}$ for $\beta = 0$ and $z = 2$ we have $\Lambda_{RCS} = \lambda_0(1 - 1/r)$, and using the numerically computed value of $\Lambda$ in the adder model yields $r = 1.887$. We find that the RCS approximation is able to better represent the behavior near the $a = l_{min}$ boundary than perturbation theory, as seen in Fig. \ref{fig:approx}.

Lastly, we can relax the RCS model by allowing the inheritance of the birth size to be a random process; namely, we let $h$ be the probability that the birth size is inherited, and with probability $1 - h$ the offspring is assigned a random birth size according to $g(l')$.   The kernel for this `Flexible Correlated Sizer' (FCS) is given by
\begin{equation}
K_{FCS}(l', l) \equiv  z e^{-\beta (2 l' - l)} \theta(l' - l_{min}) [ h \, \delta(l' - l) + (1 - h) g(l') ] \ .
\end{equation}
This kernel was previously studied in models of correlated cell division times \cite{Lebowitz:1974aa,nozoe_prl_2020}. The right eigenfunction $\psi$ has the following form, 
\begin{equation}
\psi(l') = \frac{z (1 - h) e^{ - 2 \beta l'}g(l')}{e^{\alpha}  - h z e^{ - \beta l'}} \ , \ \mathrm{for } \  l' > l_{min} \ ,
\end{equation}
and $\psi(l' \leq l_{min}) = 0$, where $\alpha$ is determined by 
\begin{equation}
\int_{l_{min}}^\infty  \frac{z (1 - h) e^{ -\beta l'} g(l')}{e^{\alpha}  - h z e^{ - \beta l'}} dl' = 1 \ .
\end{equation}
This equation has a unique solution for $\alpha$ for all $0 \leq h < 1$ \cite{nozoe_prl_2020}, and the equation can be solved numerically to yield $\alpha$ for different values of $h$ and different parameter choices $a$, $\sigma$, and $\beta$.  As in the RCS approximation, output statistics of the adder are matched by substituting $\sigma^2 \rightarrow \sigma_A^2 / 3$ in $g(l')$ above.  We note that the correlation of $l'$ and $l$ in isolated cells without mortality is equal to $h$. To determine $h$, we evaluate $\Lambda$ at $a = l_{min}$ for $\beta = 0$ and $z = 2$, which yields $\Lambda_{FCS} = \lambda_0 \log_2 (1 + h)$, and match this to the numerically computed value in the adder model, from which we obtain $h = 0.384$.  We find that the FCS approximation is slightly more accurate than RCS at intermediate values of $a$ (Fig. \ref{fig:approx}).

The approximations above show that the adder mechanism is able to adapt its lineage structure in the vicinity of mortality thresholds, which presents a major advantage over the sizer mechanism.  Biologically, this ability stems from the heritability of the birth size, which enables selection to act on the lineage structure of the population.  Physically, from the analogy of the adder mechanism with a polymer versus the sizer as a non-interacting gas, a mortality threshold is similar to a rapid change in spatial potential, e.g. near an adsorbing wall. In this case, due to the connectivity of the polymer chain, localizing one monomer to a favorable region brings neighboring monomers along, hence the per monomer entropic penalty is lower than if the monomers were a gas.

The advantage of the adder strategy in adapting to a mortality landscape is not limited to sharp thresholds.  We considered the adder and sizer models in a Gaussian mortality landscape, where survival for cells with a given birth size $l$ is given by $f(l) \equiv e^{-\frac{1}{2}\left( \frac{l - l_s}{ \sigma_s} \right)^2}$, which has peak survival at $l_s$ and width $\sigma_s$.  The kernels for sizer and adder are given by $K_S(l',l) = z g(l') f(l')$ and $K_A(l',l) = 2 z g(2l' - l) f(l')$.  In the sizer model, we obtain
\begin{equation}
\alpha_{S} =   - \frac{(a - l_s)^2}{2(\sigma^2 + \sigma_s^2)} + \ln \left( \frac{2 \sigma_s }{\sqrt{\sigma^2 + \sigma_s^2 }} \right) \label{eq:gaussian_sizer}
\end{equation}
and $\rho_S(l) \sim \mathcal{N}(\bar l, \Sigma_{S}^2)$, where $\bar l \equiv (\sigma^2 l_s + \sigma_s^2 a)/(\sigma^2 + \sigma_s^2)$ and $\Sigma_S^2 \equiv \sigma^2 \sigma_s^2 / \left(\sigma^2 + \sigma_s^2 \right)$.  In the adder model, $\alpha_{A}$ takes the same form as \eqref{eq:gaussian_sizer} and differs only in the argument of the logarithm, while $\rho_A(l) \sim \mathcal{N}(\bar l, \Sigma_{A}^2)$ with $\Sigma_{A}^2 < \Sigma_{S}^2$ \cite{supp}.  For matched output statistics, direct calculation shows that $\alpha_A > \alpha_S$, indicating the advantage of the adder mechanism in a smooth mortality landscape.

Our results are not affected by variation in the elongation rate $\lambda$, as can be seen in simulations \cite{supp}.  If we assume that the distribution of $\lambda$ is peaked around $\lambda_0$, we can obtain an analytical expression for the first-order correction to the growth rate, and show that this correction is small for realistic values of the variance in $\lambda$.  Consistent with prior results \cite{amir_cellsystems_2017,thomas2017}, this calculation shows that non-zero variance in $\lambda$ slightly reduces the long-term growth rate of a population; the small reduction is nearly identical between sizer and adder models \cite{supp}.

In summary, we have shown that in different mortality landscapes, an adder mechanism displays a long-term growth advantage over a sizer mechanism. This advantage becomes particularly pronounced near mortality thresholds, where a population using an adder mechanism can sustain positive growth while a sizer population goes extinct. Such mortality thresholds can result due to a minimal size constraint for viable cells.  In the early stages of cell cycle evolution, viability thresholds may have played a major role in natural selection acting on the control mechanism. Viability thresholds may also be important under environmental stresses in which bacterial cell sizes are reduced and therefore closer to a minimum viable size. We have shown in general that under constant growth-dependent mortality, e.g. in the presence of antibiotics, an adder mechanism is advantageous over an output-matched sizer. Occurrence of such conditions in the environment may thus select for the maintenance of the adder mechanism.  Experimental data in which both mortality and cell size are tracked in large numbers of single cells are becoming available (see e.g. \cite{nakaoka_2017}), which will enable empirical tests of our results.  We note that sizer-like behaviors can be realized experimentally via perturbations of \emph{E. coli}'s native cell division machinery \cite{Si2019-rd}, and both adders and sizers are observed in simulated network evolution of simple cell cycle models \cite{Francois2022}.  On the basis of our findings, we expect that mortality landscapes present in nature can select for and maintain the adder mechanism.

We thank T. Nozoe for helpful feedback on the manuscript.  This work was supported by NIH grant R01-GM097356.

\bibliography{cell_size_evolution_arxiv}
\clearpage
\appendix 
\renewcommand\thefigure{A\arabic{figure}}    
\setcounter{figure}{0}    
\section{Simulation and numerical methods}
\subsubsection{Simulation Method}
The simulations begin with a specified initial number of cells $N_{init}$ and a maximal population size $N_{max}$.
Each initial cell is assigned a birth size, $l$, and birth time, $t$. 
Based on the birth size and birth time, three termination values are determined, which we call the termination length, termination time, and termination type; these are represented as $l_{T}$,  $t_{T}$, and  $T$ respectively.
These terminal values for each cell in the initialized population are computed immediately, so each initial cell is represented by a five-tuple 
$c=(t,l,t_{T},l_{T},T)$.
The process to compute the three terminal values is as follows:
\begin{enumerate}
    \item A tentative division size, $l_{T}$, is computed. The simulation obtains $l_{T}$ by drawing a random value from a specified distribution.  For the Gaussian  simulations in the main text, a random value $s\sim \mathcal{N}(0,\sigma^{2})$ is drawn. 
        If the population uses the sizer mechanism, then $l_{T}=2(a+s)$.
        If the population uses the adder mechanism, then $l_{T}=l+a+s$.
        If $l_T \leq l$, then the value of $l_T$ is recomputed by drawing a new random value as above, until $l_T > l$ is obtained.
    \item Using the Inverse CDF Method, a mortality function produces a randomly drawn death size, $D$.
        If  $D \geq l_T$, then  $T =1$ (indicating a viable cell) and  $l_T$ is unchanged.
        However, if  $D<l_T$, then $l_T$ is updated to $l_T=D$, and $T=0$ (indicating a dead cell).
    \item Lastly, the simulation computes the cell's termination time using the formula $t_{T}=t+\frac{1}{\lambda}\log(l_{T}/l)$.
\end{enumerate}

Once each cell in the initialized population has the computed terminal values, the simulation updates the population using the cell with the lowest termination time. The cell with the lowest termination time dictates how the population updates based on its termination code.
If $T=0$, the cell dies, it is removed from the population, the population size decreases by 1, and the cell tuple recorded. 
If $T=1$, the cell successfully divided, the population size increases by 1, the cell tuple is recorded, and the cell is replaced by two new cells.
        The new cells are assigned a birth size of half the parent division size, and 
birth times equal to the termination time of the parent cell.  The terminal information of these new cells is then immediately computed as above.  If the maximum population size was exceeded, a randomly chosen cell is immediately removed from the population, its termination code becomes $T=-1$ (indicating removal), and its terminal size and time are updated to reflect its size at the time of removal.
The simulation then records the resulting tuple and population size.
The population updating process continues with the cell having the lowest termination time.
This process terminates when the cell with the lowest birth time is born beyond the specified total simulation time, $t_{tot}$.

In Fig. \ref{fig:intro}a, we used $N_{init} = N_{max} = 1000$ and $t_{tot} = 500$.  In Fig. \ref{fig:intro}b, we used $N_{init} = N_{max} = 50000$ and $t_{tot} = 30$, and the birth size distribution is computed from the last 500,000 birth events in the simulation. For variable $\lambda$ simulations, we used $N_{init} = N_{max} = 1000$, $t_{tot} = 500$.  

\subsubsection{Population Growth}
This simulation approach allows us to estimate the population growth rate dynamically.
As the simulation iterates through cell termination events, we track the total number of divisions, $n(t)$, minus the number of deaths, $m(t)$, to obtain a rolling total of the number of population expansive events.
This rolling total of expansive events can estimate the net rate of population growth events as $\tau^{-1} \equiv \frac{n(t)-m(t)}{t}$, where $\tau$ is the mean time between events.  Once the population reaches the maximum population size, $N_{\max}$, the population will grow from $N_{\max}$ to  $N_{\max}+1$ at each event. The population growth rate $\Lambda$ for this process is given by the relation $N_{\max}e^{\Lambda\tau}=N_{\max}+1$, or
\begin{equation}
\Lambda =\frac{n(t)-m(t)}{t} \log\left( 1+\frac{1}{N_{\max}} \right) \ .
\end{equation}
This method was used to estimate the population growth rate from a simulation.
Other methods were tested and yielded very similar results.

\subsubsection{Numerical Method}
Our numerical approach is to discretize the integral equation (Eq. \ref{eq:psi_steadystate}) into a linear equation.  We denote by $\hat K$ the operator corresponding to the kernel $K(l', l)$, such that Eq. \ref{eq:psi_steadystate} becomes $\psi =e^{-\alpha}\hat K \psi$.
We discretize this operator by forming a finite basis for the space of all possible birth size distributions, $\Phi = \{\phi_i\}$. 
To discretize the function space of birth size distributions, we assume the space of possible cell sizes is bounded, $L=[l_{\min},l_{\max}]\subset\mathbb{R}$, and we partition the space of cell sizes into $n_{bin}$ bins, each bin of width $\delta l=(l_{\max}-l_{min})/n_{bin}$.
We denote and define these bins as $l_i=[i \, \delta l,(i+1) \delta l]$ and let $\phi_i(l) =\mathbf{1}_{l_i}(l)$ denote the indicator function of a bin. We then construct a matrix $T$ as the discretization of $K$ subject to $\Phi$,
\begin{equation}
T = (T_{ij}) \equiv \int \int \phi_i(l')K(l',l)\phi_j(l)dl \, dl' \ .
\end{equation}
The eigenfunction associated with the top eigenvalue will be a linear combination of our discrete birth size basis that has the highest population growth rate.
The associated top eigenvalue will be $e^{\alpha}$.
This numerical method suggests that only modest assumptions on our choice of $K$ are needed for a normalizing choice of $\alpha$ to exist by the Perron-Frobenius Theorem.
The analog to this is the Krein-Rutman Theorem in the infinite-dimensional function space, i.e. in the limit $n_{bin} \rightarrow \infty$.

\section{Statistics of isolated single cells in adder and sizer models}
The basic size relation connecting the added size $\Delta$ and the parent-offspring birth sizes, $l'$ and $l$, is
\begin{equation}
l' = \frac{1}{2} (l + \Delta) \label{eq:basic_relation}
\end{equation}
In general terms, a sizer mechanism controls the offspring size directly, hence $l$ and $l'$ are independent random variables. An adder mechanism controls the added size directly, hence $l$ and $\Delta$ are independent random variables. This yields the simple relations between cumulants of $l$ and $\Delta$ in each model, where we denote the $n$-th cumulant of a random variable $X$ as $\kappa_n(X)$.  

We will denote by $P_l(l)$ and $P_\Delta(\Delta)$ the probability distribution function of $l$ and $\Delta$; and $M_l(\omega) \equiv \hat P_l(-\omega)$ and $M_\Delta(\omega) \equiv \hat P_\Delta(-\omega)$ are the corresponding moment generating functions, where $\hat P$ denotes the Laplace transform.  The size relation \eqref{eq:basic_relation} also yields a simple relation between the moment generating functions.

For the adder model, by independence of $l$ and $\Delta$ in Eq. \ref{eq:basic_relation}, we have $\kappa_n(l') = 2^{-n} \kappa_n(l) + 2^{-n} \kappa_n(\Delta)$. At steady state, we have $\kappa_n(l) = \kappa_n(l')$, and we obtain
\begin{equation}
\kappa_n(l) = \frac{\kappa_n(\Delta)}{2^n - 1} \ .
\end{equation}
Again by independence in Eq. \ref{eq:basic_relation}, we obtain the moment generating function relation
\begin{equation}
M_l(\omega) = M_\Delta(\omega/2)  M_l (\omega/2) \ . \label{eq:adder_MGF_relation}
\end{equation}

For the sizer model, we rewrite Eq. \ref{eq:basic_relation} as $\Delta = 2l' - l$, and by independence of $l$ and $l'$ we have $\kappa_n(\Delta) = 2^n \kappa_n(l') + (-1)^n \kappa_n(l)$.  At steady state, we obtain
\begin{equation}
\kappa_n(l) = \frac{\kappa_n(\Delta)}{2^n + (-1)^n} \ ,
\end{equation}
and by the same reasoning, we have
\begin{equation}
 M_\Delta(\omega) =  M_l(2\omega)  M_l (-\omega) \ .
\end{equation}

The control mechanisms act by determining the size added before each cell division event. The ``input'' noise into the system from each mechanism is characterized by the statistics of $\Delta$, which are fully determined by its cumulants, $\kappa_n(\Delta)$.  The ``output'' noise is characterized by the statistics of the birth size $l$.  The above equations show how the input and output noise are related in each of the models.  Given the same input noise, we see that in general
\begin{equation}
|\kappa_{n,sizer}(l)| \leq |\kappa_{n,adder}(l)| \ .
\end{equation}
Another simple consequence of the input-output relations is the distribution of $l$ or $\Delta$ for Gaussian control mechanisms, i.e. where cumulants higher than 2 are zero.  Specifically, in the adder mechanism, if $\Delta$ is Gaussian-distributed, so is $l$; while in the sizer mechanism, if $l$ is Gaussian, so is $\Delta$.  

\section{Statistics of population with simple mortality in adder and sizer models}
For cells in the presence of growth-based mortality ($\beta$) without a mortality threshold, we have $K_S(l',l) = z C_S(l') e^{-\beta(2l'-l)}$, where $C_S(l)$ is the probability distribution of the birth size $l$, and $K_A(l',l) = 2 z C_A(2l' - l) e^{-\beta(2l'-l)}$, where $C_A(\Delta)$ is the probability distribution of the added size $\Delta$. 

For the adder model, since $l, l' \geq 0$ and $C_A(\Delta) = 0$ for $\Delta < 0$, we can express Eq. \ref{eq:psi_steadystate} as 
\begin{equation}
e^{\alpha} \psi(l') = z \int_{0}^{2l'} 2 C_A(2l' - l) e^{-\beta(2l' - l)} \psi(l) dl \ .
\end{equation}
Taking the Laplace transform of both sides yields
\begin{align*}
e^\alpha &\int_{0}^{\infty} \psi(l') e^{-\omega l'} dl' = \\
&z \int_{0}^{\infty} \int_{0}^{2l'} 2 C_A(2l' - l) e^{-\beta(2l' - l)} \psi(l) e^{-\omega l'}   dl \  dl' ,
\end{align*}
and using the substitution $u = 2l'$, we have
\begin{equation}
e^\alpha \hat \psi(\omega) = z \int_{0}^{\infty} \left[ \int_{0}^{u}  C_A(u - l)e^{-\beta(u - l)} \psi(l) dl \right] e^{-\omega u/2}   du \ ,
\end{equation}
where $\hat \psi$ denotes the Laplace transform.  By the convolution theorem, we find
\begin{equation}
e^{\alpha} \hat \psi(\omega) = z \hat C_A(\beta+ \tfrac{\omega}{2}) \hat \psi (\tfrac{\omega}{2}) \ .
\end{equation}
Since $\hat \psi(0) = 1$, we obtain
\begin{equation}
\alpha_A = \ln z + \ln \hat C_A(\beta)  \ . \label{eq:alphaAgen}
\end{equation}

For the sizer model, we can express Eq. \ref{eq:psi_steadystate} as 
\begin{equation}
e^{\alpha} \psi(l') = z \int_0^\infty C_S(l') e^{-\beta(2l' - l)} \psi(l) dl \ .
\end{equation}
Taking the Laplace transform of both sides yields
\begin{equation}
e^{\alpha} \hat \psi(\omega) = z \hat C_S(\omega + 2 \beta) \hat \psi(-\beta) \ , \label{eq:psi_sizer_laplace}
\end{equation}
and letting $\omega = -\beta$ above yields
\begin{equation}
\alpha_S = \ln z + \ln \hat C_S(\beta)  \ . \label{eq:alphaSgen}
\end{equation}

To compare adder and sizer control under constant growth-based mortality, we start with a given adder control function, $C_A(\Delta)$, which has steady-state output statistics $P_l(l)$ (i.e. for isolated cells without mortality).  We compare it with the sizer control function with identical output statistics, i.e. $C_S(l) = P_l(l)$. From Eq. \ref{eq:adder_MGF_relation}, we have
\begin{equation}
 \hat C_A(\omega) = \frac{\hat C_S(2 \omega)}{\hat C_S(\omega) }  \ .
 \end{equation}
 We can now compare $\alpha_A$ and $\alpha_S$ using the formulae \eqref{eq:alphaAgen} and \eqref{eq:alphaSgen} above.  Specifically, 
 \begin{align}
 \alpha_S - \alpha_A &= \ln \hat C_S(\beta) - \ln  \frac{\hat C_S(2 \beta)}{\hat C_S(\beta) } \\
 &= 2 \ln \hat C_S(\beta) - \ln \hat C_S(2 \beta) \ .
 \end{align}
We define the function $Q(\beta) \equiv \ln \hat C_S(\beta)$, and show that it is convex by computing the second derivative as follows. 
\begin{align}
Q_S''(\beta) &= \frac{\hat C_S''(\beta)}{\hat C_S(\beta)} -  \left( \frac{\hat C_S'(\beta)}{\hat C_S(\beta)} \right)^2 \\
&= \int_0^\infty l^2 q (l) dl  - \left( \int_0^\infty l \, q (l) dl \right)^2 \ , \label{eq:convex1}
\end{align} 
where $q(l) \equiv C_S(l) e^{-\beta l} / \int_0^\infty C_S(l') e^{-\beta l'}dl'$ is a probability distribution.  Therefore, the quantity in Eq. \ref{eq:convex1} is the variance of $q(l)$ which is non-negative, yielding $Q''(\beta) \geq 0$.  Thus $Q(\beta)$ is a convex function of $\beta$, and additionally we have $Q(0) = 0$.  This implies that for any $p \in [0, 1]$, $Q(p \, x) \leq  p \, Q(x)$.  Letting $x = 2 \beta$ and $p = 1/2$, we have $2 Q(\beta) \leq  Q(2 \beta)$, which implies $\alpha_S \leq \alpha_A$.

\section{Optimal mean cell size and maximal growth rates}
To compute the optimal mean size, we take the following form, which holds for sizer exactly, and for adder in the first order approximation, 
\begin{equation}
\alpha =-  \beta a + \frac{\beta^2 \sigma^2}{2} + \ln \left[\frac{z}{2} \mathrm{erfc}(x) \right] \label{eq:alpha_general} \ .
\end{equation}
where $x = (l_{min} - a + \beta \sigma^2)/(\sigma c )$, and $c = \sqrt{2}$ for sizer or $c = \sqrt{2/3}$ for adder.  The derivative with respect to $a$ is given by
\begin{equation}
\frac{\partial \alpha}{\partial a} = -\beta +  \frac{2 e^{-x^2}}{\sigma c \sqrt{\pi} \ \mathrm{erfc}(x)} \ . \label{eq:alpha_deriv}
\end{equation}
For $a > l_{min}$ and $\sigma$ sufficiently small, we have $x \ll -1$, implying $\mathrm{erfc}(x) \approx 2$.  Setting Eq. \eqref{eq:alpha_deriv} to zero, we find
\begin{equation}
x^2 \approx -\ln(\beta \sigma c \sqrt{\pi})
\end{equation}
from which we find 
\begin{equation}
a_{opt}  \approx l_{min}  + \sigma c \sqrt{-\ln(\beta \sigma c \sqrt{\pi})} \  \label{eq:aopt1}
\end{equation}
to the lowest order in $\beta \sigma$.  We can check when is $\sigma \ll a - l_{min}$, or
\begin{equation}
\frac{a - l_{min}}{\sigma}  \approx  c \sqrt{-\ln(\beta \sigma c \sqrt{\pi})} \gg 1 \ ,
\end{equation}
which implies the condition $\sigma \ll \beta^{-1}$.  Substituting expression \eqref{eq:aopt1} in Eq. \ref{eq:alpha_general}, we find
\begin{align}
\alpha_{opt} &\approx  -\beta l_{min}  - \beta  \sigma c \sqrt{-\ln(\beta \sigma c \sqrt{\pi})}   \nonumber \\
 &\hspace{20pt}   + \ln \left[\frac{z}{2} \mathrm{erfc} \left(-\sqrt{-\ln(\beta \sigma c \sqrt{\pi})} \right)\right] \ . \label{eq:alpha_opt_approx}
\end{align}
We see that via the mapping $\sigma_S \leftrightarrow \sigma_A / \sqrt{3}$, and using $c = \sqrt{2}$ for sizer or $c = \sqrt{2/3}$ for adder, there is no difference in the values of $\alpha_{opt}$ to the lowest orders in $\beta \sigma$ computed above.  Comparison at higher orders would require higher order approximations of the adder beyond the expression given in Eq. \ref{eq:alpha_general}.
\section{Perturbation Theory}
Using the decomposition $\hat K_A = \hat K_0 + \hat K_1$, a Gaussian integral corresponds to the unperturbed operator $\hat K_0$, with kernel,
\begin{equation}
K_0(l', l) \equiv  \tfrac{2 z}{\sqrt{2\pi \sigma^2}}e^{ - \frac{(2l' - l - a)^2}{2 \sigma^2} - \beta(2l'- l) }  \ ,
\end{equation}
which yields the ground state eigenfunctions to zeroth order, and the first-order correction is due to the perturbation $\hat K_1$, with kernel
\begin{equation}
K_1(l', l) \equiv K_0(l', l) [ \theta(l' - l_{min}) - 1 ] \ .
\end{equation} 
These operators act on functions on the right or the left via integration over the kernel, e.g. 
\begin{align}
(\hat K_0 \psi)(l') &\equiv \int_{-\infty}^{\infty} K_0(l', l) \psi(l) dl \\
(\tilde \psi \hat K_0)(l) &\equiv \int_{-\infty}^{\infty} \tilde  \psi(l') K_0(l', l)  dl' \ .
\end{align}

For small noise in size control, we have $\sigma \ll a - l_{min}$, and hence $\hat K_1 \ll \hat K_0$.  We can therefore use classical perturbation theory to expand the ground state eigenvalue of $\hat K_A$, which we denote by $m$, yielding 
\begin{equation}
m = m_0 + \tilde \psi_0 \hat K_1 \psi_0 + \mathcal{O}(\hat K_1^2) \label{eq:perturbation_expansion}
\end{equation}
where $m_0$ is the leading eigenvalue of $\hat K_0$ and $\tilde \psi_0$ and $\psi_0$ are the associated left and right eigenfunctions, normalized such that $\tilde \psi_0 \psi_0 = 1$.  The $n$-th order correction to the eigenvalue will be denote $m_n$, i.e. $m_1 =  \tilde \psi_0 \hat K_1 \psi_0$.  The growth rate predicted by the $n$-th order perturbation theory is then given by $\alpha_n = \ln \left( \sum_{i = 1}^n m_i \right)$.
 
\subsubsection{Small noise limit (zeroth order)} In the lowest order approximation, we neglect the $l_{min}$ cutoff and analyze the Gaussian operator $\hat K_0$. We substitute the following functional forms
\begin{equation}
\psi_0(l) = e^{c \, l + d\, l^2}  \ \ \mathrm{and}\ \ \tilde \psi_0(l) = e^{\tilde c\, l + \tilde d\, l^2}
\end{equation}
and find two possible solutions: 
\begin{align}
\psi_0(l) &= e^{-\tfrac{3}{2 \sigma^2}[ l^2 - 2\left( a -\beta \sigma^2 \right) l ] } \label{eq:zeroth_order} \\ 
  \tilde \psi_0(l) &= 1 \nonumber \\
   m_0 &= z e^{  \tfrac{\beta^2 \sigma^2}{2} - \beta a} \nonumber
\end{align}
and
\begin{align}
  \psi_0^\prime(l) &= 1 \\
    \tilde \psi_0^\prime(l) &= e^{\tfrac{3}{2 \sigma^2}[ l^2 - 2\left( a -\beta \sigma^2 \right) l ] } \nonumber  \\
    m_0^\prime &= 2 z e^{ \tfrac{\beta^2 \sigma^2}{2} - \beta a} \nonumber
\end{align}
We can see that the solution corresponding to $m_0$ leads to a localized density by computing $\rho_0(l) \sim \psi_0(l) \tilde \psi_0(l)$, or
\begin{equation}
\rho_0(l)  = \sqrt{\tfrac{3}{2 \pi \sigma^2}} e^{-\tfrac{3}{2 \sigma^2} \left[ l - \left( a - \beta \sigma^2 \right) \right]^2 }  \ ,
\end{equation}
thus $\rho_0(l)$ has a maximum at $\bar l = a - \beta \sigma^2$, and $\rho_0(l) \rightarrow 0$ as $l \rightarrow \pm \infty$. In contrast, $\rho_0^\prime(l) \sim \psi_0^\prime(l) \tilde \psi_0^\prime(l)$ has exactly the opposite behavior, with minima and maxima interchanged. The solution corresponding to $m_0^\prime$ is therefore unphysical; it can only be realized on an unbounded space $l \in (-\infty, \infty)$. 

Thus, the physical solution is given by $\psi_0$, $\tilde \psi_0$, corresponding to a long-term growth rate $\Lambda_0 = (\lambda/\ln 2)\alpha_0$, where $\alpha_0 = \ln m_0$, or
\begin{equation}
\alpha_0 = \ln z -\beta a + \tfrac{\beta^2 \sigma^2}{2}  \ .
\end{equation}
This result takes the same form as for sizer in the $a \gg l_{min}$ limit (see Eq. \ref{eq:alpha_sizer}). The density $\rho_0(l)$ has a mean birth size $\bar l$ shifted down by $\beta \sigma^2$ relative to $a$, which reduces the effect of mortality, while its variance, $\sigma^2 / 3$, is unaffected by mortality and is the same as expected at steady state in isolated lineages of the adder model.  As mortality increases, $\bar l$ decreases, however, in the absence of constraints, $\bar l$ may decrease to negative values, with $\alpha_0$ increasing arbitrarily for large $\beta$, both of which are unphysical and result from neglecting the $l_{min}$ constraint. We must therefore require $\beta \ll a / \sigma^2$ for the above results to hold.

\subsubsection{First order perturbation}
From here on, we choose $\psi_0$ to be normalized such that $\tilde \psi_0 \psi_0 = 1$ and $\int \psi_0(l) dl = 1$, hence
\begin{align}
\psi_0(l) &= \sqrt{\tfrac{3}{2 \pi \sigma^2}} e^{-\tfrac{3}{2 \sigma^2} \left[ l - \left( a - \beta \sigma^2 \right) \right]^2 } \\ 
  \tilde \psi_0(l) &= 1 \ . \nonumber 
  \end{align}
The first order correction to $m$ is given by $m_1 \equiv \tilde \psi_0 \hat K_1 \psi_0 $, which can be evaluated as 
\begin{equation}
m_1 = z e^{\tfrac{\beta^2 \sigma^2}{2} -  \beta a} \left[ -1  +  \frac{1}{2} \mathrm{erfc} \left( \frac{l_{min} - a + \beta \sigma^2}{\sigma \sqrt{2/3}} \right) \right] \ .
\end{equation}
Together with $m_0$, we find $m \approx m_0 + m_1$ yielding
\begin{equation}
\alpha_1 \approx  -  \beta a + \tfrac{\beta^2 \sigma^2}{2} +\ln  \left[  \frac{z}{2} \mathrm{erfc} \left( \frac{l_{min} - a + \beta \sigma^2}{\sigma \sqrt{2/3}} \right) \right] \label{eq:alpha_adder}
\end{equation}
Up to here, this expression is identical to the sizer except for the factor of $\sqrt{3}$ in the argument of the complementary error function.  This factor reflects the adder's steady-state birth size variance which is $\sigma^2 / 3$.  Note that only this occurrence of $\sigma$ is scaled by $\sqrt{3}$ -- the other two occurrences involving $\beta$ are not.  We see that when $a = l_{min}$ and $\beta = 0$, we obtain $m_0 + m_1 = z/2$, thus for $z = 2$ we have $\alpha_1 = 0$ indicating that we need to go to higher orders to observe growth at the mortality threshold.

\subsubsection{Second order perturbation}
To evaluate the second order correction to $\alpha$ in general requires the full eigenbasis of the unperturbed operator $\hat K_0$.  Its left and right eigenfunctions, $\tilde v_n$ and $v_n$, for $n = 0, 1, 2, \ldots$, can be seen to take the following form:
\begin{align}
v_n &= c_n \psi_0(l) p_n(l) \\
\tilde v_n &= \tilde \psi_0(l) p_n(l) \nonumber
\end{align} 
where $p_n(l)$ is an $n$-th degree polynomial in $l$, and $c_n$ is a normalization constant.  The associated eigenvalue is $e_n \equiv 2^{-n} m_0$.  The eigenfunctions are orthonormal, i.e. $\tilde v_i v_j = \delta_{ij}$, hence the polynomials $\{ p_0(l), p_1(l), p_2(l), \ldots \}$ are orthogonal with respect to weight function $\tilde \psi_0(l) \psi_0(l)$, i.e. the density $\rho_0(l)$ which is a Gaussian.  We list the first few polynomials below:
\begin{align}
p_0(l) &= 1 , \  c_0 = 1\\
p_1(l) &= l -a +\beta \sigma^2, \  c_1 = 3 \sigma^{-2} \nonumber \\
p_2(l) &= (l - a + \beta \sigma^2)^2 -\sigma^2 / 3 , \  c_2 =(9/2) \sigma^{-4} \nonumber \\
p_3(l) &= (l - a + \beta \sigma^2)[(l - a + \beta \sigma^2)^2 - \sigma^2 ] , \  c_3 = (9/2) \sigma^{-6} \nonumber
\end{align}
The second-order correction to $m$ in Eq. \ref{eq:perturbation_expansion} is given by
\begin{equation}
m_2 = \sum_{i =1}^\infty \frac{(\tilde v_0 \hat K_1 v_i) (\tilde v_i \hat K_1 v_0)}{e_0 - e_i} \label{eq:2nd_pert_sum}
\end{equation}
For the case $\beta = 0$, the first two terms in the series yield
\begin{equation}
m_2 \approx \frac{z}{2 \pi} e^{-\tfrac{3(a - l_{min})^2}{\sigma^2}} \left[1 + \frac{(a - l_{min})^2}{2 \sigma^2} \right] \ .
\end{equation}
We see that away from the threshold (for $a > l_{min}$), $m_2$ vanishes in the limit of $\sigma \rightarrow 0$, consistent with the perturbation being small far from the threshold.  However, as $a \rightarrow l_{min}$, $m_2$ approaches a positive value which is independent of $\sigma$. In particular, exactly at the threshold (for $a = l_{min}$) the even terms in the series vanish, and evaluating up to the 5th term in Eq. \ref{eq:2nd_pert_sum} we have
\begin{equation}
m_2 \approx \frac{z}{2 \pi}\left(1 + \frac{1}{42} +\frac{3}{1240} + \ldots \right) \approx 0.163 z \ .
\end{equation}
We thus have $\alpha^{(2)} = \ln (m_0 + m_1 + m_2) \approx \ln(0.663 z)$, and for $z = 2$ we therefore obtain $\alpha^{(2)} > 0$ for $\beta = 0$ at $a = l_{min}$.  Hence, the second order perturbation theory predicts that the adder model exhibits positive growth at the mortality threshold.  In particular, the predicted long-term growth rate (setting $\lambda = 1$) is $\Lambda = \log_2(1.33) = 0.411$.  For comparison, the numerically computed growth rate at $a = l_{min}$ is $\Lambda = 0.470$.

\section{Gaussian survival function} Using the Gaussian survival function $f(l)$ defined in the main text, the sizer and adder kernels are given by
\begin{align}
K_{S}(l', l) &= \tfrac{z}{\sqrt{2 \pi \sigma^2}}e^{ - \frac{(l' - a)^2}{2 \sigma^2} } e^{-\frac{1}{2}\left( \frac{l' - l_s}{ \sigma_s} \right)^2}  \\ 
K_{A}(l', l) &= \tfrac{2 z}{\sqrt{2\pi \sigma^2}}e^{ - \frac{(2l' - l - a)^2}{2 \sigma^2}    } e^{-\frac{1}{2}\left( \frac{l' - l_s}{ \sigma_s} \right)^2}  
\end{align}

For the sizer model, we obtain
\begin{equation}
\alpha_S=   - \frac{(a - l_s)^2}{2(\sigma^2 + \sigma_s^2)} + \ln \left( \frac{z \sigma_s }{\sqrt{\sigma^2 + \sigma_s^2 }} \right) \ , \label{eq:alpha_sizer_gauss}
\end{equation}
and the population's lineage birth size distribution is
\begin{equation}
\rho_S(l) \sim \exp \left[ - \frac{1}{2 \Sigma_S^2} \left( l - \frac{\sigma^2 l_s + \sigma_s^2 a}{\sigma^2 + \sigma_s^2} \right)^2 \right]
\end{equation}
where $\Sigma_S^2 \equiv \sigma^2 \sigma_s^2 / \left(\sigma^2 + \sigma_s^2 \right)$.

In the adder model, substituting the same form for the eigenfunctions as before, we obtain analytical expressions for $\psi(l)$ and $\tilde \psi(l)$, and find
\begin{equation}
\hspace{-10pt} \alpha_A =   - \frac{(a - l_s)^2}{2(\sigma^2 + \sigma_s^2)} + \ln \left[ \frac{2 \sqrt{2} z \sigma_s }{\sqrt{\sigma^2 + 5 \sigma_s^2 + \sqrt{(\sigma^2 + \sigma_s^2)(\sigma^2 + 9 \sigma_s^2)}}} \right]  \label{eq:alpha_adder_gauss}
\end{equation}
Calculating $\rho(l) =\tilde \psi(l) \psi(l) $, we find
\[
\rho_A(l) \sim \exp \left[ - \frac{1}{2 \Sigma_A^2} \left( l - \frac{\sigma^2 l_s + \sigma_s^2 a}{\sigma^2 + \sigma_s^2} \right)^2 \right]
\]
where $\Sigma_A^2 \equiv \sigma^2 \sigma_s^2 / \sqrt{(\sigma^2 + \sigma_s^2)(\sigma^2 + 9 \sigma_s^2)}$. 

To compare adder and sizer with matched output distributions, we substitute $\sigma^2 = \sigma_S^2 = \sigma_A^2 / 3$ in Eq. \ref{eq:alpha_sizer_gauss} and $\sigma^2 = \sigma_A^2$ in Eq. \ref{eq:alpha_adder_gauss} and find
\begin{align}
\alpha_A - \alpha_S &= - \frac{(a - l_s)^2}{2(\sigma^2 + \sigma_s^2)} +  \frac{(a - l_s)^2}{2(\sigma^2/3 + \sigma_s^2)} \\
&+\frac{1}{2} \ln \left[ \frac{ 8\sigma^2/3 + 8\sigma_s^2 }{\sigma^2 + 5 \sigma_s^2 + \sqrt{(\sigma^2 + \sigma_s^2)(\sigma^2 + 9 \sigma_s^2)}} \right]  \ .
\end{align}
It is clear that the sum of the first two terms above is positive, while the logarithmic term is seen to be positive via the following calculation:
\begin{align}
&\ &8\sigma^2/3 + 8\sigma_s^2 &> \sigma^2 + 5 \sigma_s^2 + \sqrt{(\sigma^2 + \sigma_s^2)(\sigma^2 + 9 \sigma_s^2)} \nonumber \\
&\iff &(5\sigma^2 + 9\sigma_s^2)^2 &> 9(\sigma^2 + \sigma_s^2)(\sigma^2 + 9 \sigma_s^2)\nonumber \\
&\iff &16 \sigma^4 &> 0 \ .
\end{align}
This shows that $\alpha_A > \alpha_S$, indicating that in a Gaussian landscape, an adder mechanism achieves higher growth rates than a sizer with matched output.

In this model, growth rate is maximized when $a = l_s$, and there is a narrow range of sizes around $l_s$ where positive growth is possible. In the sizer model, for $|a - l_s| > \sigma_s$, there is a non-zero value of $\sigma$ that maximizes the growth rate, given by $\sigma^2 = (a - l_s)^2 - \sigma_s^2$.  In the adder model, the expression is more involved but the overall dependence is similar. In both models, the  lineage distribution of birth size is a Gaussian shifted to have a mean birth size at an intermediate value between $l_s$ and $a$; the value is determined as a linear combination weighted by the variances.  If the cell size control variance is high ($\sigma^2 \gg \sigma_s^2$), then the optimal distribution is peaked near $l_s$, and conversely if the survival variance is high, it is peaked near $a$.  Unlike the case of constant mortality $\beta > 0$ far from the mortality threshold (Eq. \ref{eq:rho_sizer} with $a \gg l_{min}$), the lineage birth size variance changes.  In the adder model, we have $\Sigma^2 < \sigma^2 / 3$, while in the sizer model we have $\Sigma^2 < \sigma^2$,   indicating that the birth size variance becomes smaller in the presence of mortality.

In the adder model, we can compute the joint density of $l$ and $\Delta$ along lineages in the population, 
\[
\rho_A(l, \Delta) \sim  \tilde \psi(\tfrac{l+\Delta}{2}) K(\tfrac{l+\Delta}{2}, l) \psi(l)   \  , 
\]
and since each term in the product is a Gaussian function, we find $\rho_A(l, \Delta)$ is a two-dimensional multivariate normal distribution; its marginal $\rho_A(l)$ was computed above.  The covariance of $\Delta$ and $l$ can be computed analytically, and will be negative, i.e. cells which are born shorter than $\bar l$ add (on average) a larger-than-average amount, and vice versa.  Thus, due to selection, lineages in the population deviate from the adder-like behavior of isolated cells in which $\Delta$ and $l$ are independent. 

\section{Variability of $\lambda$}
We start with Eq. \ref{eq:B_steadystate} of the main text, 
\begin{equation}
B(l') =  \int  (2l' / l)^{-\Lambda / \lambda}  K(l', l)   B(l) p(\lambda)   \, dl \, d\lambda \ .
\end{equation}
We can define the $\lambda$-averaged kernel,
\begin{align}
\hat L = L(l', l) &\equiv  \int  (2l' / l)^{-\Lambda / \lambda}   K(l', l)   p(\lambda) d\lambda 
\end{align}
and seek the (positive) right and left eigenfunctions of $\hat L$, $B(l)$ and $\tilde B(l)$, which we normalize such that $\int B(l) dl = 1$ and $\int \tilde B(l) B(l) dl = 1$.  Then the equation for $\Lambda$ is given by setting the corresponding eigenvalue of $\hat L$, which is the multiplicative growth rate of the statistical weight, equal to 1, thus normalizing the right eigenfunction:
\begin{equation}
\bra{\tilde B} \hat L \ket{B}  =  \int \tilde B(l') L(l', l) B(l) dl dl' = 1
\end{equation}

We have already solved the case of $p(\lambda) = \delta(\lambda - \lambda_0)$, i.e. fixed $\lambda = \lambda_0$, corresponding to the kernel
\begin{equation}
\hat L_0 =  L_0(l', l) \equiv (2l' / l)^{-\Lambda_0 / \lambda_0} K(l', l)  
\end{equation}
where we  obtain $B_0(l)$ and $\tilde B_0(l)$, the right and left eigenfunctions of $\hat L_0$, which are related to the eigenfunctions $\psi$ and $\tilde \psi$ of $\hat K$ via $B_0(l) = l^{-\Lambda_0/\lambda_0} \psi(l)$ and $\tilde B_0(l) = l^{\Lambda_0/\lambda_0} \tilde \psi(l)$; these are normalized as above, thus $\int \tilde \psi(l) \psi(l) dl = 1$ and $\int l^{-\Lambda_0/\lambda_0} \psi(l) dl = 1$.  

Generalizing now for the case of $p(\lambda)$ which is strongly peaked around $\lambda = \lambda_0$ (e.g. a Gaussian with mean $\lambda_0$ and width $\sigma_\lambda \ll \lambda_0$), we can expand the operator $\hat L$ around $\hat L_0$, and we treat $\hat L - \hat L_0$ as a small perturbation of $\hat L_0$.  From perturbation theory, we know that the perturbed eigenvalue is given by $\bra{B_0} \hat L \ket{B_0}$, so we can in principle directly calculate the eigenvalue by evaluating the integral, given that we have already calculated the ground-state eigenfunctions from the case $\lambda = \lambda_0$. We expand the operator $\hat L$ around $\lambda_0$ in a Taylor series to the second order, 
\begin{widetext}
\begin{equation}
L(l', l) \approx (2 l'/  l)^{-\frac{\Lambda}{\lambda_0}} K(l', l)  \left[ 1 + \frac{\Lambda}{\lambda_0^2} \log (2 l'/  l) \int \left( \lambda - \lambda_0 \right)p(\lambda) d\lambda \right. 
\left.  +\frac{1}{2} \frac{\Lambda}{\lambda_0^3}  \log (2 l'/  l)  \left( \frac{\Lambda}{\lambda_0}   \log (2 l'/  l) - 2  \right) \int (\lambda - \lambda_0)^2 p(\lambda) d\lambda \right] \ .
\end{equation}
\end{widetext}
The first order term above is zero since the expectation of $\lambda$ is $\lambda_0$, thus the first-order perturbation of the operator is given by the second term,
\begin{align}
L_1&(l', l) \equiv (2 l'/  l)^{-\frac{\Lambda }{\lambda_0}} K(l', l) \\
&\ \times  \frac{\sigma_\lambda^2}{ 2\lambda_0^{2}}\frac{\Lambda  }{\lambda_0}  \log (2 l'/  l)  \left( \frac{\Lambda}{\lambda_0}   \log (2 l'/  l) - 2  \right) \ ,
\end{align}
where $\sigma^2_\lambda$ is the variance of $p(\lambda)$.  
A hand-waiving argument to show $\bra{\tilde B_0} \hat L_1 \ket{B_0} < 0$ is that we expect $\Lambda / \lambda_0 \leq 1$, and due to size control we have on average that $l' \approx l$, therefore the value in the parentheses is less than approximately $\log (2) - 2$, which is negative, and if $\Lambda > 0$ this implies $\bra{\tilde B_0} \hat L_1 \ket{B_0} < 0$.

For the case of the sizer with a gamma-distributed birth size distribution and no mortality, we can compute things explicitly.  We use the kernel
\begin{equation}
K(l', l) = z \, {l'}^{k - 1} \theta^{-k} e^{-l' / \theta} / \Gamma(k)
\label{eq:gamma_kernel}
\end{equation}
which has shape parameter $k$ and scale parameter $\theta$.  In the absence of mortality, we have $\Lambda_0 = \lambda_0$ and we find the ground-state eigenfunctions
\begin{align}
B_0(l) &= l^{-2 + k} \theta^{1 -k } e^{-l / \theta} / \Gamma(k - 1) \nonumber \\
\tilde B_0(l) &= l \theta^{-1}  / (k - 1)
\end{align}
where the eigenfunctions are normalized as needed; this requires $k > 1$.  Now, we do not need to expand the operator $\hat L$ first, we can directly evaluate the perturbed eigenvalue to first order as
\begin{equation}
\bra{B_0} \hat L \ket{B_0}= \int d\lambda \frac{2^{- \Lambda/\lambda} z \, p(\lambda) }{\Gamma(k)^2} \Gamma \left( k + \tfrac{\Lambda}{\lambda} - 1 \right)  \Gamma \left( k - \tfrac{\Lambda}{\lambda} + 1 \right). \label{eq:numeric_gamma}
\end{equation}
In the limit of a high shape parameter $k$, the integrand approaches $2^{-\Lambda/\lambda}z \, p(\lambda)$. Expanding around $\lambda = \lambda_0$ to second order, we have 
\begin{equation}
\bra{B_0} \hat L \ket{B_0} \approx 2^{-\Lambda/\lambda_0} z \left[1+ \tfrac{\Lambda}{\lambda_0} \tfrac{\sigma_\lambda^2}{\lambda_0^2} \ln 2 \left(-1 + \tfrac{\Lambda}{\lambda_0} \tfrac{ \ln 2}{2} \right) \right]
\end{equation}
To find $\Lambda$, we set the above equal to 1, and solve for $\Lambda$.  Expanding $\Lambda$ around $\lambda_0$, we obtain for $\Lambda - \lambda_0 \ll 1$ and $\sigma_\lambda /\lambda_0 \ll 1$
\begin{equation}
\frac{1}{2} +\frac{\sigma_\lambda^2 \ln 2 (\ln 2 - 2)}{4 \lambda_0^2} - \frac{\ln 2}{2 \lambda_0} (\Lambda - \lambda_0)  = \frac{1}{z}
\end{equation} 
For $z = 2$, we obtain the solution
\begin{equation}
\Lambda = \lambda_0 -\frac{\sigma_\lambda^2}{\lambda_0} \left( 1 - \frac{\ln 2}{2}  \right)  \ .
\label{eq:approx_gamma}
\end{equation}
Thus, we recapitulate the known result for the Gaussian model \cite{amir_cellsystems_2017}, which shows how a small variance in $\lambda$ reduces the long-term growth rate.

\begin{figure*}
\includegraphics[width=7in]{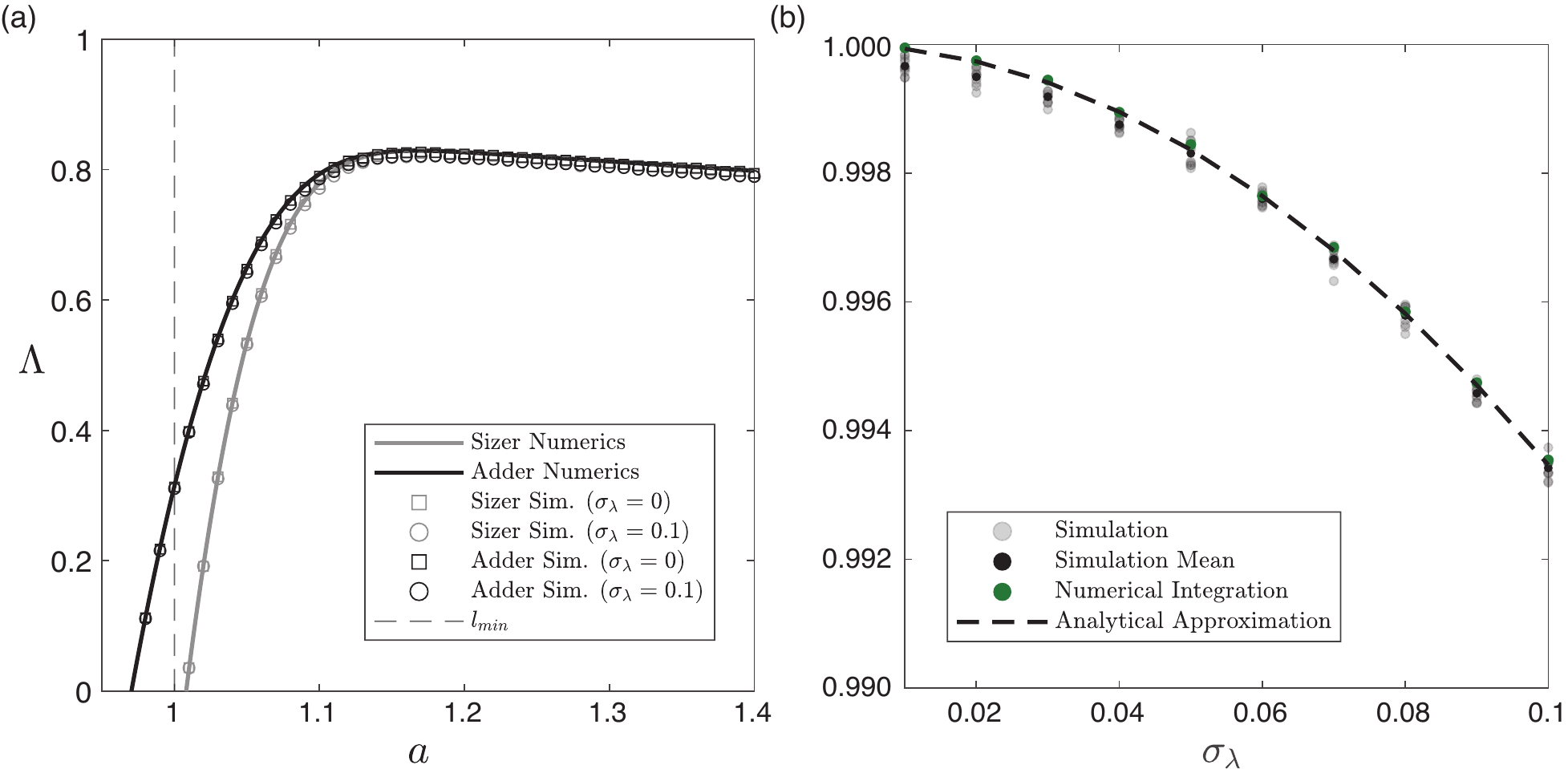}
\caption{Effect of variation of elongation rates on long-term growth rate. (a) Comparison of growth rates in sizer and adder models with variable elongation rate.  Points indicate simulation results with indicated parameter values.  Curves show prediction from transfer operator for constant elongation rate (Eq. \ref{eq:psi_steadystate}). Parameter values are $\sigma_A = 0.1$, $\sigma_S = \sigma_A / \sqrt{3}$, $\beta = 0.1$, $l_{min} = 1$, and $\lambda_0 = 1$; the choice of $\sigma_S$ ensures that sizer and adder have identical output statistics in isolated cells. (b) Simulation and analytical results for the gamma-distributed sizer kernel of Eq. \ref{eq:gamma_kernel} using variable elongation rates that are normally distributed with mean $\lambda_0$ and variance $\sigma_\lambda^2$. The plot shows $\Lambda$ at different values of $\sigma_\lambda$ using $k = 100$, $\theta = 0.01$, and $\lambda_0 = 1$.  Gray points indicate simulation results, with black points showing the mean over simulations.  Green points are the numerical prediction from Eq. \ref{eq:numeric_gamma} and the dashed curve shows the analytical approximation in Eq. \ref{eq:approx_gamma}.  }
\label{fig:variation_of_lambda}
\end{figure*}
In Fig. \ref{fig:variation_of_lambda}, we show that theoretical prediction based on numerical evaluation of Eq. \ref{eq:numeric_gamma} closely matches simulation results, and that the approximate analytical result in Eq. \ref{eq:approx_gamma} likewise captures the overall trend of decreases $\Lambda$ with increases $\sigma_\lambda^2$.  In general, the effect of $\lambda$ variation is seen to be quite small, reducing the population growth rate by less than 1\% for experimentally realistic values of $\sigma_\lambda$.

\begin{figure*}
\includegraphics[width=7in]{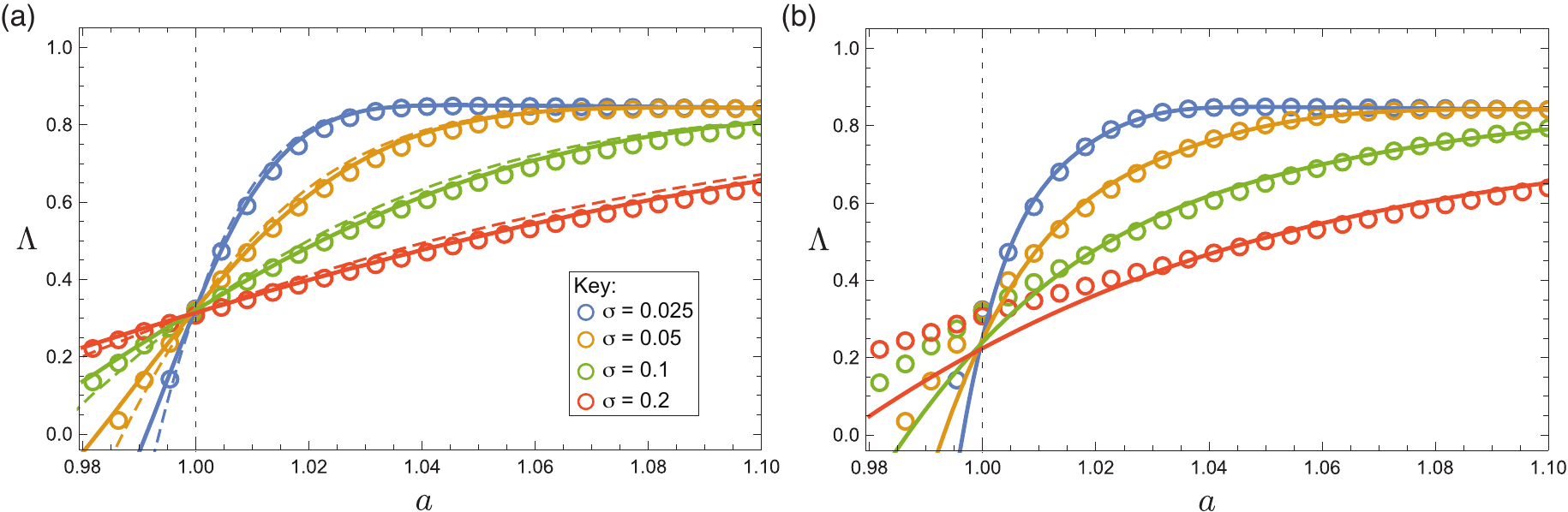}
\caption{Analytical results in the adder model for $\beta = 0.1$.  Numerically computed values (points) and analytical approximations (curves) are shown for the long-term growth rate, $\Lambda$, as a function of the mean birth size, $a$. (a) RCS approximation, $r = 1.887$ (dashed lines) and FCS approximation, $h = 0.384$ (solid lines). Parameter values $\sigma$ are indicated in the key; additionally, $\beta = 0.1$, $\l_{min} = 1$, $z = 2$, and $\lambda_0 = 1$. (b) Perturbation theory, second order (solid lines).}
\label{fig:suppfig2}
\end{figure*}

\end{document}